\documentclass[prb,twocolumn,longbibliography,superscriptaddress,nofootinbib]{revtex4-1}

\usepackage[utf8]{inputenc}
\usepackage[T1]{fontenc}
\usepackage{amsmath}
\usepackage{amssymb}
\usepackage{amsthm}
\usepackage{siunitx}
\usepackage{physics}
\usepackage{graphicx}
\usepackage{subfigure}
\usepackage{tabularx}
\usepackage{booktabs}
\usepackage{multirow}
\usepackage[pdftex,colorlinks=true,urlcolor=blue,linkcolor=blue,citecolor=blue]{hyperref}        
\usepackage{xcolor}

\usepackage{soul}
\usepackage{ulem}

\newcolumntype{Y}{>{\centering\arraybackslash}X}
\newcolumntype{P}[1]{>{\centering\arraybackslash}p{#1}}

\begin{document}

\title{Many-Body Quadrupolar Sum Rule for Higher-Order Topological Insulator}

\author{Wonjun Lee}
\thanks{Electronic Address: wonjun1998@postech.ac.kr}
\affiliation{Department of Physics, Pohang University of Science and Technology (POSTECH), Pohang 37673, Republic of Korea}
\affiliation{Center for Artificial Low Dimensional Electronic Systems, Institute for Basic Science (IBS), Pohang 37673, Republic of Korea}

\author{Gil Young Cho}
\thanks{Electronic Address: gilyoungcho@postech.ac.kr}
\affiliation{Department of Physics, Pohang University of Science and Technology (POSTECH), Pohang 37673, Republic of Korea}
\affiliation{Center for Artificial Low Dimensional Electronic Systems, Institute for Basic Science (IBS), Pohang 37673, Republic of Korea}
\affiliation{Asia Pacific Center for Theoretical Physics, Pohang 37673, Republic of Korea}

\author{Byungmin Kang}
\thanks{Electronic Address: bkang119@kias.re.kr}
\affiliation{School of Physics, Korea Institute for Advanced Study, Seoul 02455, Korea}

\date{\today}

\begin{abstract} 
The modern theory of polarization establishes the bulk-boundary correspondence for the bulk polarization. In this paper, we attempt to extend it to a sum rule of the bulk quadrupole moment by employing a many-body operator introduced in [\href{https://doi.org/10.1103/PhysRevB.100.245134}{Phys. Rev. B 100, 245134 (2019)}] and [\href{https://doi.org/10.1103/PhysRevB.100.245135}{Phys. Rev. B 100, 245135 (2019)}]. The sum rule that we propose consists of the alternating sum of four observables, which are the phase factors of the many-body operator in different boundary conditions. We demonstrate its validity through extensive numerical computations for various non-interacting tight-binding models. We also observe that individual terms in the sum rule correspond to the bulk quadrupole moment, the edge-localized polarizations, and the corner charge in the thermodynamic limit on some models.
\end{abstract}

\maketitle

\section{Introduction}
Recent developments in topological insulators~\cite{RevModPhys.82.3045, RevModPhys.83.1057} discovered a new class of topological material, called higher-order topological insulators~\cite{benalcazar2017quantized, Benalcazar2017, schindler2018higher}. Higher-order topological insulators are characterized by non-trivial boundary-of-boundary states despite that the boundaries are trivial. Among various theoretical tools, the bulk polarization has proven useful in understanding many aspects of topological insulators~\cite{PhysRevB.74.195312, PhysRevLett.102.107603, PhysRevB.83.245132, PhysRevB.84.075119, PhysRevB.89.155114}. It is therefore natural to ask if the bulk quadrupole or higher multipole moments could play the same role for higher-order topological insulators.

The modern theory of polarization~\cite{PhysRevB.47.1651, resta1993macroscopic, PhysRevB.48.4442} identifies the bulk polarization with the sum of Wannier centers. Later, the bulk polarization of a many-body state is shown to be identified as the phase factor of the expectation value of a many-body operator~\cite{Resta1998}. Despite the success of the modern theory of polarization, the modern theory of quadrupole and higher multipole moments has not been fully developed so far. One attempt is to employ the Wannier centers to define the bulk quadrupole moment, but this approach has been successful when there exist crystalline symmetries so that the bulk multipole moment is quantized~\cite{Benalcazar2017, PhysRevB.102.165120}. On the other line of development, Refs.~\onlinecite{Kang2019, Wheeler2019} introduced a many-body operator $\hat{U}_2$ for the quadrupole moment generalizing the many-body operator presented in Ref.~\onlinecite{Resta1998}. This approach can in principle be applied to systems without any crystalline symmetries and to interacting systems, however, the observables given by $\hat{U}_2$ have difficulties in identifying their physical meanings with respect to their coordinate dependence~\cite{Kang2019} and lack of periodicity~\cite{PhysRevB.100.245133} so their physical meaning have not been fully understood yet.

One of the success of the modern theory of polarization is that it identifies the bulk polarization with the boundary charge, which is called the bulk-boundary correspondence~\cite{PhysRevB.48.4442, Rhim2017}. While the bulk polarization is determined by a full many-body electron state, the boundary charge is determined by a simple one-body observable near the boundary. Thus, the identification of the two is rather surprising and also provides a firm justification of the modern theory of polarization. A natural question is then whether such a correspondence for the bulk quadrupole can be formulated. In recent studies~\cite{PhysRevResearch.2.043012, PhysRevB.103.035147}, bulk-boundary correspondences for the quadrupole moment were presented using Wannier functions, where the corner charge is expressed as a sum of quantities that depend on the choice of the bulk Wannier functions, although the corner charge itself is not. We also note another recent study~\cite{ying2021change}, where the corner charge of band insulators is computed using the adiabatic current flowing along the edges.

In this paper, we propose another bulk-boundary correspondence in terms of the operator $\hat{U}_2$, which is given by an alternating sum of four phase factors, where the sum vanishes in the thermodynamic limit:
\begin{equation*}
	\phi_{pp}-\phi_{op}-\phi_{po}+\phi_{oo} = 0 \quad\text{(mod 1)}.
\end{equation*}
Four phase factors are the phase factors of the expectation values of $\hat{U}_2$ with respect to the ground state in different boundary conditions,
\begin{equation*}
	\phi_{ab} = \frac{1}{2\pi}\text{Im} \Big[ \log \big( \langle \textrm{GS}(a,b) \vert \hat{U}_2 \vert \textrm{GS}(a,b) \rangle \big) \Big],
\end{equation*}
where `$a$' and `$b$' refer to as the boundary conditions along $x$- and $y$-direction, respectively, with `$a$' and `$b$' being either `$p$' (periodic) or `$o$' (open). We observe that for band insulators having well-localized edge-localized polarizations defined in Ref.~\onlinecite{Benalcazar2017} based on the hybrid Wannier function (HWF), the phase factors $\phi_{pp}$, $\phi_{po/op}$, and $\phi_{oo}$ can be identified with the bulk quadrupole moment, the edge-localized polarizations, and the corner charge, respectively, which consequently elevates our bulk-boundary correspondence to that for the quadrupole moment. We further observe that for some insulators, there exists HWFs with the corresponding hybrid Wannier value at $0.5$ so that one cannot directly apply the definition of the edge-localized polarization presented in Ref.~\onlinecite{Benalcazar2017} due to its branch cut dependence, as elaborated in Appendix~\ref{appendix:edge-pol}. For those insulators, we find that at least the phase factor $\phi_{oo}$ can be regarded as the corner charge in the thermodynamic limit while the physical meaning of the other phase factors becomes unclear.

The paper is organized as follows. In Sec.~\ref{sec:sum-rules}, we introduce sum rules encoding the bulk-boundary correspondence for multipole moments using many-body operators. In Sec.~\ref{sec:numerics}, we provide numerical results presenting our observations in Sec.~\ref{sec:sum-rules}. We conclude in Sec.~\ref{sec:conclusion} where the summary of main results and possible future directions are discussed. In Appendix~\ref{appendix:classical_multipole_sum_rule}, we derive the dipolar and quadrupolar sum rules for classical systems. In Appendix~\ref{appendix:U1}, we provide a proof that the many-body operator can measure the corner charge in one-dimensional geometry for band insulators. In Appendix~\ref{appendix:C3-model}, we provide numerical results for a $C_3$-symmetric insulator, which suggests that the sum rule may hold in other types of symmetric insulators besides the $C_4$-symmetric ones discussed in the main text. Finally, in Appendix~\ref{appendix:edge-pol}, we provide technical remarks on the HWF-based edge-localized polarization including the branch cut dependence which becomes more explicit for the models with the hybrid Wannier value at $0.5$.

\section{Multipolar Sum Rules} \label{sec:sum-rules}
In the following, we present the multipolar sum rules for quantum systems in one- and two-dimension. For concreteness, we note here some remarks on the sum rules. First, we only focus on one-dimensional systems having non-zero bulk polarization in Sec.~\ref{sec:dipole} and two-dimensional systems with vanishing bulk polarization in Sec.~\ref{sec:quadrupole}. Second, we focus on circle/line geometry for one-dimensional systems and torus/cylinder/rectangle geometry for two-dimensional systems. Each geometry is characterized by open or periodic boundary conditions. In addition, we always use hard-cutoff boundary conditions for open boundaries. Third, for simplicity we restrict the position of each orbital at a Bravais lattice site. Fourth, for all the quantum systems presented below, we assume translation symmetry and $C_2$ symmetry, which includes the cases where the corner charge is not quantized. Finally, for convenience, we set the electron charge to 1 in the following discussions.

\subsection{Dipolar Sum Rule}\label{sec:dipole}
In this subsection, we introduce the dipolar sum rule for many-body quantum systems as a generalization of the classical dipolar sum rule. The dipolar sum rule relates the bulk polarization with the boundary charge. To motivate the quantum mechanical dipolar sum rule, we first state the classical dipolar sum rule,
\begin{equation}\label{eq:classical_dipole_sum_rule}
	\mathcal{Q}_c = \mathcal{P},
\end{equation}
where $\mathcal{P}$ and $\mathcal{Q}_c$ are the classical bulk polarization and boundary charge, respectively. The derivation of Eq.~\eqref{eq:classical_dipole_sum_rule} and the definitions of the electric moments can be found in Appendix~\ref{appendix:classical_dipole_sum_rule}.

We now generalize the classical dipolar sum rule Eq.~\eqref{eq:classical_dipole_sum_rule} to a one-dimensional quantum mechanical system. To this end, we first need to define the bulk polarization and the boundary charge for a quantum system. The boundary charge can be defined as a one-body observable:
\begin{equation}\label{eq:one_dim_cc}
	Q^{(1)}_c\equiv \sum_{x=1}^{L_x/2} \big(\rho(x) - n_e \big) \quad\text{(mod 1)},
\end{equation}
where $L_x$ is the total system size, $\rho(x) = \langle \hat{n}_x \rangle = \langle \sum_{\alpha=1}^{N_\textrm{orb}} c_{x, \alpha}^\dagger c_{x, \alpha} \rangle$ is the charge density at site $x$ with the number of orbitals per site $N_{\rm orb}$, and $n_e$ is the average number of electrons per site. We then employ the Resta's many-body operator~\cite{Resta1998}
\begin{equation}
	\hat{U}_1 \equiv \exp(\frac{2\pi i}{L_x}\sum_{x=1}^{L_x} x (\hat{n}_x - n_e) ) ,
\end{equation}
where we include the background charge $n_e$ from ions sitting at each lattice site. Using $\hat{U}_1$, the bulk polarization is given by
\begin{equation}
	P = \frac{1}{2\pi} \Im \Big[ \log \big( \langle \textrm{GS} \vert \hat{U}_1 \vert \textrm{GS} \rangle \big) \Big],
\label{eq:U1-pol}
\end{equation}
where $\vert \textrm{GS} \rangle$ is the many-body ground state subject to the periodic boundary condition. With these, the classical dipolar sum rule Eq.~\eqref{eq:classical_dipole_sum_rule} generalizes as~\cite{PhysRevB.48.4442, Rhim2017}
\begin{equation}\label{eq:dipole_sum_rule}
Q^{(1)}_c = P\quad\text{(mod 1)} 
\end{equation}
in quantum systems. Unlike the dipole sum rule in classical systems, Eq.~\eqref{eq:dipole_sum_rule} holds only modulo 1, the unit of electron charge. 

Having obtained the sum rule for the dipole moment, we consider the further characterization solely in terms of the many-body operator $\hat{U}_1$. For this, we define the phase factor $\phi_a$ of the expectation value of $\hat{U}_1$ as 
\begin{equation}
	\phi_a \equiv \frac{1}{2\pi}\text{Im} \Big[ \log \big( \langle \textrm{GS}(a) \vert \hat{U}_1 \vert \textrm{GS}(a) \rangle \big) \Big],
\end{equation}
where $\ket{\textrm{GS}(a)}$ is the ground state under the boundary conditions of type `$a$' which can be either `$p$' (periodic) or `$o$' (open).

While it seems not widely known, the boundary charge can also be captured from $\phi_o$ in the thermodynamic limit, i.e.,
\begin{equation}\label{eq:cc_phio}
	Q^{(1)}_c = \phi_o\quad\text{(mod 1)} 
\end{equation}
as the system size goes to infinity when the system is gapped. Our proof of Eq.~\eqref{eq:cc_phio} for band insulators can be found in Appendix~\ref{appendix:U1}. Thus, by combining Eqs.~(\ref{eq:U1-pol}) and (\ref{eq:cc_phio}), the dipole sum rule can be succinctly recast in terms of the phase factors:
\begin{equation}\label{eq:phi_dipole_sum_rule}
	\phi_p= \phi_o \quad\text{(mod 1)}.
\end{equation}

\subsection{Quadrupolar Sum Rule}
\label{sec:quadrupole}
In this subsection, we introduce the quadrupolar sum rule for many-body quantum systems, which generalizes the quantum mechanical dipolar sum rule to the quadrupole case, and discuss its difficulties for certain cases in identifying the physical meaning of individual terms in the sum rule.

The quadrupole sum rule would relate the bulk quadrupole moment, the edge-localized polarization, and the corner charge~\cite{Benalcazar2017, Kang2019} if these quantities are well-defined. To illustrate the quadrupolar sum rule, we first need to state the classical quadrupolar sum rule,
\begin{equation}\label{main-eq:classical_quad_sum_rule}
	\mathcal{Q}_c = -\mathcal{Q}_{xy} + \mathcal{P}^{\text{edge}}_{x} + \mathcal{P}^{\text{edge}}_{y},
\end{equation}
where $\mathcal{Q}_{xy}$, $\mathcal{P}^{\text{edge}}_{x/y}$, and $\mathcal{Q}_c$ are the classical bulk-quadrupole moment, the edge-localized polarizations, and the corner charge, respectively. The derivation of Eq.~\eqref{main-eq:classical_quad_sum_rule} and the definitions of the classical electric moments can be found in Appendix~\ref{appendix:classical_quadrupole_sum_rule}.

A natural generalization of the classical quadrupolar sum rule to the quantum mechanical one in two dimensions would be 
\begin{equation}\label{eq:quad_sum_rule}
	Q^{(2)}_c = -Q_{xy} + P^{\text{edge}}_{x} + P^{\text{edge}}_{y}\quad\text{(mod 1)},
\end{equation}
where $Q^{(2)}_c$ is the corner charge of a two dimensional system of the size $L_x\times L_y$,
\begin{equation}\label{eq:two_dim_cc}
	Q^{(2)}_c \equiv \sum_{x=1}^{L_x/2}\sum_{y=1}^{L_y/2} \big( \rho(x,y) - n_e \big) \quad\text{(mod 1)},
\end{equation}
with $\rho(x,y) = \langle \hat{n}_{x,y} \rangle = \langle \sum_{\alpha=1}^{N_{\rm orb}} c_{x, y, \alpha}^\dagger c_{x,y, \alpha} \rangle$ being the local charge density, $Q_{xy}$ and $P^{\text{edge}}_{x/y}$ are the quantum mechanical bulk quadrupole moment and the edge-localized polarizations, respectively. As a side note, our corner charge Eq.~\eqref{eq:two_dim_cc} corresponds to the bare corner charge in Ref.~\onlinecite{PhysRevB.103.035147}. We remark that the edge-localized polarizations $P^{\text{edge}}_{x/y}$ include the contribution from a dressing of polarized one-dimensional chains along the boundaries~\cite{PhysRevB.100.054408}, which does not affect the quadrupolar sum rule Eq.~\eqref{eq:quad_sum_rule}, and are fixed for a given state. The precise forms of $Q_{xy}$ and $P^{\text{edge}}_{x/y}$ will be presented below.

Similar to the dipole case, let us introduce the following many-body operator~\cite{Kang2019,Wheeler2019}:
\begin{equation}\label{eq:U2}
	\hat{U}_2 \equiv \exp(\frac{2\pi i}{L_xL_y}\sum_{x,y=1}^{L_x,L_y}xy (\hat{n}_{x,y} - n_e)),
\end{equation}
where $\hat{n}_{x,y}$ is the electron number operator at $(x,y)$ and $L_x$ ($L_y$) is the linear system size in $x$-direction ($y$-direction). For the open boundary, we assign the coordinate $x=1$ ($y=1$) to sites on the left (bottom) boundary and the coordinate $x=L_x$ ($y=L_y$) to sites on the right (top) boundary. For the periodic boundary, the $x$-coordinate ($y$-coordinate) could be assigned arbitrary as long as it starts from $1$ and ends with $L_x$ ($L_y$) since it does not change the expectation value of $\hat{U}_2$ with respect to a translation invariant state.

In order to formulate the quadrupolar sum rule in terms of the phase factors of the expectation values of $\hat{U}_2$, we first introduce
\begin{equation} \label{eq:U2_phase}
	\phi_{ab} = \frac{1}{2\pi}\text{Im} \Big[ \log \big( \langle \textrm{GS}(a,b) \vert \hat{U}_2 \vert \textrm{GS}(a,b) \rangle \big) \Big],
\end{equation}
where $\ket{\textrm{GS}(a,b)}$ is the ground state under the `$a$' and `$b$' boundary conditions along $x$- and $y$-direction, respectively, with `$a$' and `$b$' being either `$p$' (periodic) or `$o$' (open). Each boundary condition corresponds to different geometry, for example, `$pp$' and `$po$' mean the torus and the cylinder geometry, respectively. These phase factors are gauge independent by construction. 

Using these phase factors, we propose the bulk-boundary correspondence in terms of the phase factors as
\begin{equation}\label{eq:phi_quad_sum_rule}
	\phi_{pp}-\phi_{op}-\phi_{po}+\phi_{oo} = 0 \quad\text{(mod 1)}
\end{equation}
and numerically find that this holds in the thermodynamic limit on band insulators. We will call this as the sum rule of the phase factors, or simply the sum rule. In addition, we also find that $\phi_{oo}$ agrees with $Q^{(2)}_c$,
\begin{equation}\label{eq:phi_oo}
	Q^{(2)}_c = \phi_{oo} \quad\text{(mod 1)}.
\end{equation}
As a demonstration, we provide numerical confirmations of these on band insulators in Sec.~\ref{sec:numerics}. The above two equations, i.e., Eqs.~\eqref{eq:phi_quad_sum_rule} and~\eqref{eq:phi_oo}, are the main results of our present paper.

Based on the identification of the corner charge Eq.~\eqref{eq:phi_oo} and the comparison between two sum rules Eqs.~\eqref{eq:quad_sum_rule} and \eqref{eq:phi_quad_sum_rule} suggest that $\phi_{pp}$ can be naturally identified with the bulk quadrupole moment
\begin{equation}\label{eq:phi_pp}
	Q_{xy} = \phi_{pp} \quad\text{(mod 1)},
\end{equation}
which has been checked for the cases with non trivial bulk quadrupole moment~\cite{Kang2019, Wheeler2019}. In addition, $\phi_{po/op}$ are identified with the edge-localized polarizations
\begin{equation}\label{eq:phi_op_po}
	\begin{split}
		P^{\text{edge}}_{x} &= \phi_{po} \quad\text{(mod 1)} \\
		P^{\text{edge}}_{y} &= \phi_{op} \quad\text{(mod 1)}.
 	\end{split}
\end{equation}
Whenever these identifications can be made, the sum rule Eq.~\eqref{eq:phi_quad_sum_rule} indeed becomes the quadrupole sum rule for quantum systems. In Sec.~\ref{sec:with_edgepol}, we numerically demonstrate the validity of the identifications of $\phi_{po/op}$ by comparing these with the HWF-based edge-localized polarizations defined in Ref.~\onlinecite{Benalcazar2017}. However, it is important to note that these identifications are not always possible. In Sec.~\ref{sec:without_edgepol}, we provide an example on which these identifications fail as $\phi_{pp}$, $\phi_{op}$, and $\phi_{po}$ are not convergent in the thermodynamic limit. Nonetheless, even for these cases, the sum rule Eq.~\eqref{eq:phi_quad_sum_rule} and the identification of the corner charge with $\phi_{oo}$ Eq.~\eqref{eq:phi_oo} always hold.

In the remainder of this section, we shortly discuss the HWF-based edge-localized polarization\cite{Benalcazar2017},
\begin{equation}\label{eq:edgepol}
	\tilde{P}^{\textrm{edge}}_x = \sum_j \sum_{y=1}^{L_y/2}\nu^j \rho^j(y)\quad\text{(mod 1)},
\end{equation}
where $\rho^j(y)$ is the density, and $e^{2\pi i\nu_j}$ is the eigenvalue of the $j$-th HWF $\psi^j(y)$ of the hybrid Wilson loop $\mathcal{W}_{k_x}$ along the $x$-direction. We find that for band insulators having well-localized $\tilde{P}^{\textrm{edge}}_{x/y}$ along the boundaries, $\tilde{P}^{\textrm{edge}}_{x/y}$ and $\phi_{po/op}$ seem to agree each other in the thermodynamic limit and the same quantized value for insulators having $C_4$ symmetry,
\begin{equation}\label{eq:phi_edgepol}
	\begin{split}
		\tilde{P}^{\textrm{edge}}_x &= \phi_{op} \quad\text{(mod 1)}\\
		\tilde{P}^{\textrm{edge}}_y &= \phi_{po} \quad\text{(mod 1)}.
	\end{split}
\end{equation}
However, $\tilde{P}^\textrm{edge}_{x/y}$ crucially depends on the choice of the branch cut of the hybrid Wannier values. In particular, for the models with the hybrid Wannier value at $0.5$, this depdendance becomes more explicit, thereby leading to a difficulty in computing $\tilde{P}_{x/y}^\textrm{edge}$. We discuss this branch cut dependence more in details in Appendix~\ref{appendix:edge-pol}.

\section{Numerical Demonstration} \label{sec:numerics}
In this section, we provide numerical demonstrations of the sum rule Eq.~\eqref{eq:phi_quad_sum_rule} and the observations, Eqs.~\eqref{eq:phi_oo} and \eqref{eq:phi_edgepol}. Due to size limitation, our numerics are based on non-interacting tight-binding models on the square lattice. However, we believe that the same sum rule should also hold in interacting cases as well, since our formalism is based on many-body operators. In addition, we separately discuss models without and with the hybrid Wannier value at $0.5$ as there exists a difficulty in computing $\tilde{P}^\textrm{edge}_{x/y}$ for the latter case. We therefore numerically confirm the validity of Eq.~\eqref{eq:phi_edgepol} only for the former case while the numerical confirmation of the sum rule Eq.~\eqref{eq:phi_quad_sum_rule} and Eq.~\eqref{eq:phi_oo} are presented in all cases. 

\subsection{Models without the hybrid Wannier value at $0.5$}\label{sec:with_edgepol}
We present numerical results on models without the hybrid Wannier value at $0.5$. Tested models are the quadrupole insulator, the edge-localized polarization insulator, the quadrupole insulator with $\pi/2$-flux per plaquette, and the two band model introduced in Ref.~\onlinecite{PhysRevB.92.041102}, where details of these models can be found below. In the case of full open boundary conditions, $C_4$ symmetry breaking term is always introduced in order to split the possible degeneracy at the Fermi level. Due to such term, the number of filled states is always equal to the filling $n_e/N_\textrm{orb}$ times the number of sites $N$. The same $C_4$ breaking term often splits the degeneracy of the hybrid Wannier values at $0.5$ in the case of mixed open and periodic boundary conditions.

\begin{figure*}[t]
	\centering\includegraphics[width=0.95\textwidth]{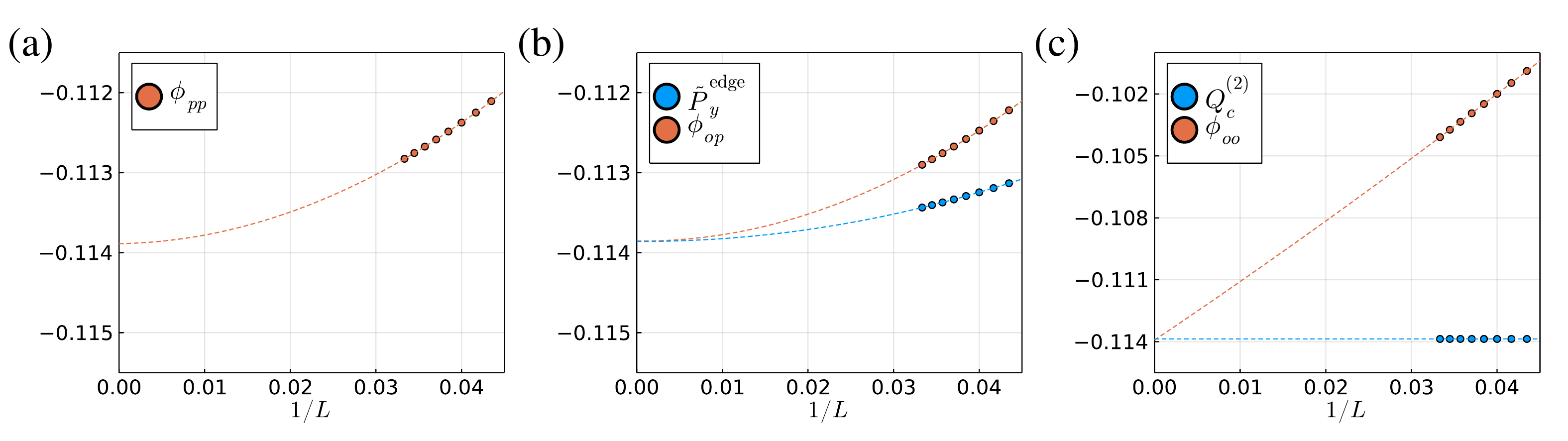}
	\caption{Extrapolation of the observables (a) $\phi_{pp}$, (b) $\phi_{op}$ and $\tilde{P}^{\text{edge}}_y$, (c) $\phi_{oo}$ and $Q^{(2)}_c$, as a quadratic function in $1/L$ for the quadrupole insulator Eq.~\eqref{eq:ham-quad} with parameters $(\gamma_x,\gamma_y,\lambda_x,\lambda_y,\delta)=(0.5,0.5,1.0,1.0,0.7)$ and isotropic system sizes.	The quadratic extrapolation works well and the error from the extrapolation is less than $10^{-5}$ in all cases. (b) $\phi_{op}$ and $\tilde{P}^{\text{edge}}_y$ agree with each other in $O(10^{-5})$. (c) $\phi_{oo}$ and $Q^{(2)}_c$ agree with each other in the same order of accuracy.
	}
	\label{Fig:fitting}
\end{figure*}

For each model, we numerically compute the phase factors $\phi_{ab}$ Eq.~\eqref{eq:U2_phase}, the edge-localized polarizations $\tilde{P}_{x/y}^\textrm{edge}$ Eq.~\eqref{eq:edgepol}, and the corner charge $Q_c^{(2)}$ Eq.~\eqref{eq:two_dim_cc} for various parameters. The results are summarized in Tables~\ref{table:quad-insulator},~\ref{table:edge-insulator}, and~\ref{table:quad-pi/2-insulator}. All observables in the tables are the extrapolated values in the thermodynamic limit $L \to \infty$ which are obtained via quadratic extrapolations as a function of $1/L$, as shown in Fig.~\ref{Fig:fitting}. In all cases, we find that the phase factors in the sum rule and the sum rule itself are convergent in the thermodynamic limit. We also see that our edge-localized polarizations, $\phi_{op/po}$, have values similar to the other ones, $\tilde{P}_{x/y}^\textrm{edge}$, with differences at worst $O(10^{-3})$. In addition, when a $C_4$ symmetry breaking term is small, $\phi_{pp}$ has the same quantized value as the one predicted by the nested Wilson-loop approach~\cite{Benalcazar2017}. Furthermore, the identification of $\phi_{oo}$ with $Q^{(2)}_c$ and the sum rule hold up to errors of $O(10^{-4})$. The errors may be attributed to the finite-size effect.

\subsubsection{Quadrupole insulator $H_{\text{quad}}$}
As our first model with gapped Wannier spectrum, we consider the following tight-binding model~\cite{Benalcazar2017} having a non-zero bulk quadrupole moment:
\begin{align} \label{eq:ham-quad}
H_\textrm{quad} (\boldsymbol{k}) = &\big( \gamma_x + \lambda_x \cos(k_x) \big) \Gamma_4 + \lambda_x \sin(k_x) \Gamma_3 \\
&+ \big( \gamma_y + \lambda_y \cos(k_y) \big) \Gamma_2 + \lambda_y \sin(k_y) \Gamma_1 + \delta \, \Gamma_0 \nonumber ,
\end{align}
where $\Gamma_0 = \sigma_3 \otimes \sigma_0$, $\Gamma_k = - \sigma_2 \otimes \sigma_k$ for $k=1,2,3$, and $\Gamma_4 = \sigma_1 \otimes \sigma_0$ with $\sigma_{k=1,2,3}$ being Pauli matrices and $\sigma_0$ being the $2 \times 2$ identity matrix. 

When $\delta=0$, there exists two anti-commuting mirror symmetries which quantize the bulk quadrupole moment $Q_{xy}=0$ or $1/2$. The half-filled ground state of Eq.~\eqref{eq:ham-quad} realizes topologically trivial quadrupole insulator when $|\gamma_x| > |\lambda_x|$ and $|\gamma_y| > |\lambda_y|$ and topologically non-trivial quadrupole insulator when $|\gamma_x| < |\lambda_x|$ and $|\gamma_y| < |\lambda_y|$. When $\delta \ne 0$, $C_4$ symmetry and the mirror symmetries are broken and hence the bulk quadrupole moment is no longer quantized. 

The numerical results for the quadrupole insulator are summarized in Table~\ref{table:quad-insulator}.

\begin{table*}
	\footnotesize
	\centering
	\begin{ruledtabular}
	\renewcommand{\arraystretch}{1.4}
	\begin{tabular}{rrrrrrrrr}
		\multicolumn{1}{c}{Model parameters}  & \multicolumn{4}{c}{Phase factors} & \multicolumn{3}{c}{Electric moments} & \multicolumn{1}{c}{Sum rule} \\
		\cmidrule{1-1}\cmidrule{2-5}\cmidrule{6-8}\cmidrule{9-9}
		$H_{\text{quad}}(\gamma_x,\gamma_y,\lambda_x,\lambda_y,\delta)$ & $\phi_{pp}$ & $\phi_{po}$ & $\phi_{op}$ & $\phi_{oo}$ & $Q^{(2)}_c$ & $\tilde{P}_x^\textrm{edge}$ & $\tilde{P}_y^\textrm{edge}$ & $\sum (-1)^{ab} \phi_{ab}$\\
		\hline
		$(0.1,0.1,1.0,1.0,10^{-7})$	
		& 0.50000(0) & 0.50000(0) & 0.50000(0) & 0.50000(0) & 0.50000(0) & 0.50000(0) & 0.50000(0) & 0.00000(0)\\
		\hline
		$(0.5,0.5,1.0,1.0,0.7)$ 
		& -0.11389(0) & -0.11385(0) & -0.11385(0) & -0.11392(0) & -0.11387(0) & -0.11386(0) & -0.11386(0) & -0.00010(0)\\
		\hline
		$(0.2,0.3,1.0,1.0,0.1)$
		& -0.43156(0) & -0.43155(0) & -0.43155(0) & -0.43155(0) & -0.43156(0) & -0.43156(0) & -0.43155(0) & -0.00001(0)\\
		\hline
		$(0.1,0.1,1.0,1.2,0.1)$ 	
		& -0.44018(0) & -0.44017(0) & -0.44018(0) & -0.44019(0) & -0.44018(0) & -0.44018(0) & -0.44018(0) & -0.00002(0)\\
		\end{tabular}
	\end{ruledtabular}
	\caption{
		The phase factors $\phi_{ab}$ in Eq.~\eqref{eq:U2_phase} and the electric moments $(Q_c^{(2)},\tilde{P}_x^{\text{edge}},\tilde{P}_y^{\text{edge}})$ in Eqs.~\eqref{eq:two_dim_cc} and \eqref{eq:edgepol} are computed for the quadrupole insulator Eq.~\eqref{eq:ham-quad} with various parameters. The sum of the phase factors $\sum (-1)^{ab} \phi_{ab}$ in the last column refers to the combination $\phi_{pp} - \phi_{po} -\phi_{op} + \phi_{oo}$ as per Eq.~\eqref{eq:phi_quad_sum_rule}. All values are the ones from extrapolating the observables as a function of $1/L$, where we consider isotropic systems $L_x=L_y=L$ with $L$ from 23 to 30. An explicit extrapolation procedure is presented in Fig.~\ref{Fig:fitting}. We use round brackets to denote the errors of the least significant digit. We see that the edge-localized polarizations $\tilde{P}_x^\textrm{edge}$ and $\tilde{P}_y^\textrm{edge}$ agree with phase factors $\phi_{po}$ and $\phi_{op}$, corner charge $Q^{(2)}_c$ agrees with $\phi_{oo}$, and the sum rule is satisfied up to small errors.
	}
	\label{table:quad-insulator}
\end{table*}

\subsubsection{Edge-localized polarization insulator $H_{\text{edge}}$}
Our next model is the edge-localized polarization insulator~\cite{Benalcazar2017}. The tight-binding Hamiltonian in the momentum space can be written as
\begin{align} \label{eq:ham-edge-pol}
H_{\text{edge}}(\boldsymbol{k}) = &\big( \gamma_x + \lambda_x \cos(k_x) \big) \Gamma_4 + \lambda_x \sin(k_x) \Gamma_3 \\
&+ \big( \gamma_y + \lambda_y \cos(k_y) \big) \tilde{\Gamma}_2 + \lambda_y \sin(k_y) \tilde{\Gamma}_1 + \delta \, \Gamma_0 \nonumber ,
\end{align}
where $\tilde{\Gamma}_2 = \sigma_1 \otimes \sigma_1$, $\tilde{\Gamma}_1 = -\sigma_1 \otimes \sigma_2$, and all the other $\Gamma$ matrices are the same as the ones in Eq.~\eqref{eq:ham-quad}.

The half-filled ground state of Eq.~\eqref{eq:ham-edge-pol} has a vanishing bulk quadrupole moment but having a non-zero corner charge. In particular, when $\delta=0$, there exist two mirror symmetries $M_x$ and $M_y$ which quantize the corner charge to be $0$ or $1/2$, where the corner charge is originated sorely from the edge-localized polarizations~\cite{Benalcazar2017}. The half-filled ground state of Eq.~\eqref{eq:ham-edge-pol} with $\abs{\gamma_{x/y}} \ll \abs{\lambda_{x/y}}$ has $(\tilde{P}_x^{\textrm{edge}},\tilde{P}_y^{\textrm{edge}})=(0, 0.5)$ when $\lambda_x > \lambda_y$ and $(\tilde{P}_x^{\textrm{edge}},\tilde{P}_y^{\textrm{edge}})=(0.5, 0)$ when $\lambda_x < \lambda_y$.

The numerical results for the edge-localized polarization insulator are summarized in Table~\ref{table:edge-insulator}.

\begin{table*}
	\footnotesize
	\centering
	\begin{ruledtabular}
	\renewcommand{\arraystretch}{1.4}
	\begin{tabular}{rrrrrrrrr}
		\multicolumn{1}{c}{Model parameters} & \multicolumn{4}{c}{Phase factors} & \multicolumn{3}{c}{Electric moments} & \multicolumn{1}{c}{Sum rule} \\
		\cmidrule{1-1}\cmidrule{2-5}\cmidrule{6-8}\cmidrule{9-9}
		$H_{\text{edge}}(\gamma_x,\gamma_y,\lambda_x,\lambda_y,\delta)$ & $\phi_{pp}$ & $\phi_{po}$ & $\phi_{op}$ & $\phi_{oo}$ & $Q^{(2)}_c$ & $\tilde{P}_x^\textrm{edge}$ & $\tilde{P}_y^\textrm{edge}$ & $\sum (-1)^{ab} \phi_{ab}$\\
		\hline
		$(0.1,0.1,1,1.2,10^{-7})$ 	
		& 0.00000(0) & 0.50000(0) & 0.00000(0) & 0.50000(0) & 0.50000(0) & 0.50000(0) & 0.00000(0) & 0.00000(0)\\
		\hline
		$(0.5,0.5,1.0,1.0,0.7)$ 	
		& -0.01661(0) & -0.09914(0) & -0.09914(0) & -0.18179(0) & -0.18174(0) & -0.09836(0) & -0.09836(0) & -0.00012(0)\\
		\hline
		$(0.2,0.3,1.0,1.0,0.1)$
		& -0.05940(1) & -0.27871(1) & -0.31926(1) & 0.46139(1) & 0.46122(1) & -0.27833(1) & -0.31897(1) & -0.00004(1) \\
		\hline
		$(0.1,0.1,1.0,1.2,0.1)$
		& -0.01585(0) & -0.44086(0) & -0.12064(0) & 0.45436(0) & 0.45434(0) & -0.44172(0) & -0.11936(0) & 0.00002(0) \\
		\end{tabular}
	\end{ruledtabular}
	\caption{
	The phase factors $\phi_{ab}$ and the electric moments $(Q_c^{(2)},\tilde{P}_x^{\text{edge}},\tilde{P}_y^{\text{edge}})$ 
	with are computed for the edge-localized polarization insulator Eq.~\eqref{eq:ham-edge-pol} with various parameters. Here we use the same extrapolation procedure as in Table.~\ref{table:quad-insulator}. In this case as well, the edge-localized polarizations $\tilde{P}_x^\textrm{edge}$ and $\tilde{P}_y^\textrm{edge}$ agree with phase factors $\phi_{po}$ and $\phi_{op}$, corner charge $Q^{(2)}_c$ agrees with $\phi_{oo}$, and the sum rule is satisfied up to small errors.
	}
	\label{table:edge-insulator}
\end{table*}

\subsubsection{Quadrupole insulator with $\pi/2$-flux per plaquette $H_{\text{quad}}^{(\pi/2)}$}
Our third model is the quadrupole insulator with $\pi/2$-flux per plaquette~\cite{Kang2019}. The tight-binding Hamlitonian in momentum space can be written as
\begin{align} \label{eq:ham-quad-pi/2}
	H_{\text{quad}}^{(\pi/2)} (\boldsymbol{k}) = 
	&\gamma_x\Gamma_4+\gamma_y(i\nu_2\otimes\nu_2) +\delta\Gamma_0\\
	&-\lambda_x(\cos(k_x)(\sigma_y\otimes\sigma_0)+\sin(k_x)(\sigma_x\otimes\sigma_z))\nonumber\\
	&-\lambda_y(\cos(k_y)(\nu_1\otimes\nu_2)+\sin(k_y)(i\nu_1\otimes\nu_1))\nonumber,
\end{align}
where $\nu_1$ and $\nu_2$ are defined as
\begin{equation}
	\nu_1 =
	\begin{bmatrix}
		0 & 1 \\
		i & 0
	\end{bmatrix}
	\quad\text{and}\quad
	\nu_2 =
	\begin{bmatrix}
		0 & 1 \\
		-i & 0
	\end{bmatrix},
\end{equation}
and all $\Gamma$ and $\sigma$ matrices are the same as in Eq.~\eqref{eq:ham-quad}.

\begin{table*}
	\footnotesize
	\centering
	\begin{ruledtabular}
	\renewcommand{\arraystretch}{1.4}
	\begin{tabular}{rrrrrrrrr}
		\multicolumn{1}{c}{Model parameters} & \multicolumn{4}{c}{Phase factors} & \multicolumn{3}{c}{Electric moments} & \multicolumn{1}{c}{Sum rule} \\
		\cmidrule{1-1}\cmidrule{2-5}\cmidrule{6-8}\cmidrule{9-9}
		$H_{\text{quad}}^{(\pi/2)}(\gamma_x,\gamma_y,\lambda_x,\lambda_y,\delta)$ & $\phi_{pp}$ & $\phi_{po}$ & $\phi_{op}$ & $\phi_{oo}$ & $Q^{(2)}_c$ & $\tilde{P}_x^\textrm{edge}$ & $\tilde{P}_y^\textrm{edge}$ & $\sum (-1)^{ab} \phi_{ab}$\\
		\hline
		$(0.1,0.1,1.0,1.0,10^{-7})$	
		& 0.50000(0) & 0.50000(0) & 0.50000(0) & 0.50000(0) & 0.50000(0) & 0.50000(0) & 0.50000(0) & 0.00000(0)\\
		\hline
		$(0.5,0.5,1.0,1.0,0.7)$ 	
		& -0.06272(0) & -0.09622(0) & -0.09622(0) & -0.12980(0) & -0.12975(0) & -0.09675(0) & -0.09675(0) & -0.00009(0)\\
		\hline
		$(0.2,0.3,1.0,1.0,0.1)$
		& -0.40237(0) & -0.42169(0) & -0.42104(0) & -0.44037(0) & -0.44039(0) & -0.42192(0) & -0.42155(0) & -0.00001(0)\\
		\hline
		$(0.1,0.1,1.0,1.2,0.1)$ 	
		& -0.42101(0) & -0.43853(0) & -0.43160(0) & -0.44913(0) & -0.44913(0) & -0.43855(0) & -0.43165(0) & -0.00001(0)\\
		\end{tabular}
	\end{ruledtabular}
	\caption{
		The phase factors $\phi_{ab}$ and the electric moments $(Q_c^{(2)},\tilde{P}_x^{\text{edge}},\tilde{P}_y^{\text{edge}})$ are computed for the quadrupole insulator with $\pi/2$-flux per plaquette Eq.~\eqref{eq:ham-quad-pi/2} with various parameters. Here we use the same extrapolation procedure as in Table.~\ref{table:quad-insulator}. In this case as well, the edge-localized polarizations $\tilde{P}_x^\textrm{edge}$ and $\tilde{P}_y^\textrm{edge}$ agree with phase factors $\phi_{po}$ and $\phi_{op}$, corner charge $Q^{(2)}_c$ agrees with $\phi_{oo}$, and the sum rule is satisfied up to small errors.
	}
	\label{table:quad-pi/2-insulator}
\end{table*}
Similar to the quadrupole insulator, which has $\pi$-flux per plaquette, the bulk quadrupole moment of the half-filled ground state of Eq.~\eqref{eq:ham-quad-pi/2} is quantized when $\delta = 0$. However, when $\delta \ne 0$, the bulk quadrupole moment $Q_{xy}$ is non-zero and $(Q_{xy}, \tilde{P}_{x/y}^\textrm{edge}, Q^{(2)}_c)$ are {\it all} distinct unlike in the quadrupole insulator. These features provide non-trivial checks of the sum rule Eq.~\eqref{eq:phi_quad_sum_rule} and our observations Eqs.~\eqref{eq:phi_oo} and \eqref{eq:phi_edgepol}, which are summarized in Table~\ref{table:quad-pi/2-insulator}.

\subsubsection{Two band insulator $H_{two}$}
Our final model is the two band insulator introduced in Ref.~\onlinecite{PhysRevB.92.041102}. The tight-binding Hamlitonian in momentum space can be written as
\begin{align} \label{eq:ham-two-band}
	H_{\text{two}} (\boldsymbol{k}) = \frac{t_1}{2}\sigma_1 + 
	e^{ik_x}\begin{bmatrix}
		t_4 & t_2 \\
		0 & t_4
	\end{bmatrix}+
	e^{ik_y}\begin{bmatrix}
		t_5 & t_3 \\
		0 & t_5
	\end{bmatrix}+h.c.,
\end{align}
with the five hopping parameters $(t_1, t_2, t_3, t_4, t_5)$. The half filled ground state of this Hamiltonian has non-zero corner charge and edge-localized polarizations. 

The numerical results for the two band insulator with the same parameters as in Ref.~\onlinecite{PhysRevB.92.041102} are summarized in Table~\ref{table:two-band-insulator}.

\begin{table*}
	\footnotesize
	\centering
	\begin{ruledtabular}
	\renewcommand{\arraystretch}{1.4}
	\begin{tabular}{rrrrrrrrr}
		\multicolumn{1}{c}{Model parameters} & \multicolumn{4}{c}{Phase factors ($\times 10^{-6}$)} & \multicolumn{3}{c}{Electric moments ($\times 10^{-6}$)} & \multicolumn{1}{c}{Sum rule ($\times 10^{-6}$)} \\
		\cmidrule{1-1}\cmidrule{2-5}\cmidrule{6-8}\cmidrule{9-9}
		$H_{\text{two}}(t_1,t_2,t_3,t_4,t_5)$ & $\phi_{pp}$ & $\phi_{po}$ & $\phi_{op}$ & $\phi_{oo}$ & $Q^{(2)}_c$ & $\tilde{P}_x^\textrm{edge}$ & $\tilde{P}_y^\textrm{edge}$ & $\sum (-1)^{ab} \phi_{ab}$\\
		\hline
		\multirow{1}{*}{$(-2.2,-0.15,-0.1,-0.09,-0.06)$}
		& 0.000(0) & 1.459(0) & 1.459(0) & 2.920(0)	& 2.918(0) & 0.851(0) & 0.851(0) & 0.002(0)
		\end{tabular}
	\end{ruledtabular}
	\caption{
		The phase factors $\phi_{ab}$ and the electric moments $(Q_c^{(2)},\tilde{P}_x^{\text{edge}},\tilde{P}_y^{\text{edge}})$ are computed for the two band insulator $H_{\text{two}}$ Eq.~\eqref{eq:ham-two-band}. Here we use the same extrapolation procedure as in Table.~\ref{table:quad-insulator}. In this case as well, the edge-localized polarizations $\tilde{P}_x^\textrm{edge}$ and $\tilde{P}_y^\textrm{edge}$ agree with phase factors $\phi_{po}$ and $\phi_{op}$, corner charge $Q^{(2)}_c$ agrees with $\phi_{oo}$, and the sum rule is satisfied up to small errors.
	}
	\label{table:two-band-insulator}
\end{table*}

\subsection{Models with the hybrid Wannier value at $0.5$}\label{sec:without_edgepol}
We now turn to models with the hybrid Wannier value at $0.5$, which are all based on $C_4$-symmetric insulators~\cite{Benalcazar2019}. Note that for the models presented below, a HWF with the corresponding hybrid Wannier value at $0.5$ exists even after introducing a $C_4$ symmetry breaking term. In these models, corner charges are originated from the filling anomaly $\eta$~\cite{Benalcazar2019}:
\begin{equation}
\eta = \textrm{\# ions} - \textrm{\# electrons} \quad \textrm{mod} \,\, 4 ,
\end{equation}
where the number of ions equals the number of lattice sites times the filling. In each model, we take into account the filling anomaly $\eta$ which determines the corner charge and the phase factor $\phi_{oo}$, where we tune the chemical potential to include or not include the zero modes in the full open boundary conditions. 

For tested models which are specified below, we numerically compute the phase factors $\phi_{ab}$ Eq.~\eqref{eq:U2_phase} and the corner charge $Q_c^{(2)}$ Eq.~\eqref{eq:two_dim_cc}. The results are summarized in Tables~\ref{table:C42b-insulator} and \ref{table:C41b+2c-insulator}. In the tables, $\tilde{P}_{x/y}^{\textrm{edge}}$ are not presented since they depend sensitively on the choice of the branch cut for the considered models, as we discuss in Appendix~\ref{appendix:edge-pol}. However, we do compare the corner charge $Q^{(2)}_c$ with $\phi_{oo}$ and check whether the sum rule Eq.~\eqref{eq:phi_quad_sum_rule} is satisfied. All observables in the tables are the extrapolated values in the thermodynamic limit as described in Sec.~\ref{sec:with_edgepol}. 

For each model, we separate the cases based on the quadrupole moment $\phi_0$ of the background ions,
\begin{equation}\label{eq:ion_quadpol}
	\phi_0 \equiv -n_e(L_x+1)(L_y+1)/4 \quad\text{(mod 1)},
\end{equation}
with $n_e$ being the charge of the ions at each lattice site and $(L_x,L_y)$ being the linear system sizes. We then obtain $\phi_{ab}$ and $Q_c^{(2)}$ in the thermodynamic limit $L\rightarrow\infty$ for each value of $\phi_{0}$ and summarize the numerical results in Tables~\ref{table:C42b-insulator} and \ref{table:C41b+2c-insulator}. We find that for the model Eq.~\eqref{eq:ham-C4-1b-2c} defined below, the convergent values of the three phase factors $\phi_{pp}$, $\phi_{op}$, and $\phi_{po}$ depend on $\phi_0$ while the phase factor $\phi_{oo}$ is convergent on the same model. One the contrary, all the phase factors are convergent on the other model Eq.~\eqref{eq:ham-2b}. Thus, at least $\phi_{oo}$ is interpretable among the four phase factors for these models. In addition, we find that in all cases, $\phi_{oo}$ converges to corner charge $Q^{(2)}_c$ up to an error $O(10^{-5})$, and the sum rule holds up to the same order of an error.

\subsubsection{$C_4$-symmetric insulator $H_{2b}^{(4)}$}
We first consider a $C_4$-symmetric insulator $H_{2b}^{(4)}$ introduced in Ref.~\onlinecite{Benalcazar2019} with a $C_4$ symmetry breaking term parameterized by $\delta$:
\begin{equation} \label{eq:ham-2b}
	H^{(4)}_{2b}(\boldsymbol{k}) = 
	\begin{bmatrix}
		\delta & t & e^{i(k_x+k_y)} & t \\
		t & -\delta & t & e^{-i(k_x-k_y)} \\
		e^{-i(k_x+k_y)} & t & \delta & t \\
		t & e^{i(k_x-k_y)} & t & -\delta
	\end{bmatrix} .
\end{equation}
When $\delta=0$ and $t < 1$, the half-filled ground state of Eq.~\eqref{eq:ham-2b} respects $C_4$ symmetry and has filling anomaly $\eta = 2$. Therefore on full open boundary conditions, filling $2N - 2$ electrons, where $N$ is the number of sites, results in the corner charge of $-1/2$ (in unit of electron charge) at each corner. When $\delta \ne 0$, the ground state no longer respects $C_4$ symmetry and the corner charge is not quantized.

The numerical results for the $C_4$-symmetric insulator $H_{2b}^{(4)}$ are summarized in Table~\ref{table:C42b-insulator}.

\begin{table*}
	\footnotesize
	\centering
	\begin{ruledtabular}
	\renewcommand{\arraystretch}{1.4}
	\begin{tabular}{rrrrrrrr}
		\multicolumn{1}{c}{Model parameters} & \multicolumn{1}{c}{Background} & \multicolumn{4}{c}{Phase factors} & \multicolumn{1}{c}{Electric moment} & \multicolumn{1}{c}{Sum rule} \\
		\cmidrule{1-1}\cmidrule{2-2}\cmidrule{3-6}\cmidrule{7-7}\cmidrule{8-8}
		$H^{(4)}_{2b}(t,\delta)$ & $\phi_0$ & $\phi_{pp}$ & $\phi_{po}$ & $\phi_{op}$ & $\phi_{oo}$ & $Q^{(2)}_c$ & $\sum (-1)^{ab} \phi_{ab}$\\
		\hline
		\multirow{2}{*}{$(0.1,0.001)$} 
		& 0.5 & \multirow{2}{*}{0.50000(0)} & \multirow{2}{*}{0.49992(0)} & \multirow{2}{*}{0.49992(0)} & \multirow{2}{*}{0.49983(0)} & \multirow{2}{*}{0.49983(0)} & \multirow{2}{*}{0.00000(0)} \\
		& 0 & & & & & & \\
		\hline
		\multirow{2}{*}{$(0.1,0.1)$} 
		& 0.5 & 0.49972(0) & 0.49433(0) & 0.49433(0) & \multirow{2}{*}{0.48896(0)} & \multirow{2}{*}{0.48895(0)}  & \multirow{2}{*}{0.00002(0)} \\
		& 0 & 0.49976(0) & 0.49435(0) & 0.49435(0) & & & \\
		\hline
		\multirow{2}{*}{$(0.1,0.5)$} 
		& 0.5 & 0.49732(2) & 0.49095(1) & 0.49095(1) & \multirow{2}{*}{0.48461(0)} & \multirow{2}{*}{0.48460(0)} & 0.00003(0) \\
		& 0 & 0.49732(0) & 0.49095(0) & 0.49095(0) & & & 0.00002(0) \\
		\end{tabular}
	\end{ruledtabular}
	\caption{
		The phase factors $\phi_{ab}$ and the corner charge $Q_c^{(2)}$	are computed for $H^{(4)}_{2b}$ in Eq.~\eqref{eq:ham-2b} with various parameters. The number of filled electrons is taken as $2N-2$ with the number of sites $N$ when the system is under the full open boundary condition due to the filling anomaly $\eta=2$. Here we use the same extrapolation procedure as in Table.~\ref{table:quad-insulator}. The corner charge agrees with $\phi_{oo}$ and the sum rule Eq.~\eqref{eq:phi_quad_sum_rule} is satisfied up to small errors.
	}
	\label{table:C42b-insulator}
\end{table*}

\subsubsection{$C_4$-symmetric insulator $H^{(4)}_{1b}\oplus H^{(4)}_{2c}$} 
\label{subsubsection:C4-1b+2c}
Finally, we consider the $C_4$-symmetric insulator~\cite{Benalcazar2019} $H^{(4)}_{1b}\oplus H^{(4)}_{2c}$. This model was used as an example on which $\phi_{pp}$ depends on the system size~\cite{PhysRevB.100.245133}, which indicates it is ill-defined on this model. We also see that $\phi_{pp}$ in fact does as discussed below. The tight-binding Hamiltonian of this model in the momentum space is given by
\begin{equation}\label{eq:ham-C4-1b-2c}
	H^{(4)}_{1b}\oplus H^{(4)}_{2c}(\boldsymbol{k}) = 
	\begin{bmatrix}
		H^{(4)}_{1b} & \gamma^{(4)}(t) \\ 
		\gamma^{(4)}(t)^\dagger & H^{(4)}_{2c}
	\end{bmatrix} ,
\end{equation}
where
\begin{equation}
	H^{(4)}_{1b}(\boldsymbol{k}) = 
	\begin{bmatrix}
		0 & t_x+e^{ik_x} & 0 & t_y+e^{ik_y}\\
		t_x+e^{-ik_x} & 0 & t_y+e^{ik_y} & 0\\
		0 & t_y+e^{-ik_y} & 0 & t_x+e^{-ik_x}\\
		t_y+e^{-ik_y} & 0 & t_x+e^{k_x} & 0
	\end{bmatrix},
\end{equation}
\begin{equation}
	H^{(4)}_{2c}(\boldsymbol{k}) = 
	\begin{bmatrix}
		0 & 0 & t_x & 0\\
		0 & 0 & 0 & t_y\\
		t_x & 0 & 0 & 0\\
		0 & t_y & 0 & 0
	\end{bmatrix}
	+ 1.5
	\begin{bmatrix}
		0 & 0 & e^{ik_x} & 0\\
		0 & 0 & 0 & e^{ik_y}\\
		e^{-ik_x} & 0 & 0 & 0\\
		0 & e^{-ik_y} & 0 & 0
	\end{bmatrix},
\end{equation}
and
\begin{equation}
	\gamma^{(4)}(t) = 
	\begin{bmatrix}
		t & t & 0 & 0\\
		0 & t & t & 0\\
		0 & 0 & t & t\\
		t & 0 & 0 & t
	\end{bmatrix} .
\end{equation}
Both the $1/4$-filled ground state of $H^{(4)}_{1b}$ and the $1/2$-filled ground state of $H^{(4)}_{2c}$ have non-zero bulk polarizations, and we stack two models and introduce additional hopping term given by $\gamma^{(4)} (t)$ so that the $3/8$-filled ground state of the stacked model has vanishing bulk polarization while respecting $C_4$ symmetry. The stacked model $H^{(4)}_{1b}\oplus H^{(4)}_{2c}$ has the filling anomaly $\eta = 3$ which results in $-3/4$ corner charge (in the unit of electron charge) at each corner when $4N - 3$ electrons are filled under full open boundary conditions, where $N$ is the number of sites.

The numerical results for the $C_4$-symmetric insulator $H^{(4)}_{1b}\oplus H^{(4)}_{2c}$ are summarized in Table~\ref{table:C41b+2c-insulator}. 

\begin{table*}
	\footnotesize
	\centering
	\begin{ruledtabular}
	\renewcommand{\arraystretch}{1.4}
	\begin{tabular}{rrrrrrrr}
		\multicolumn{1}{c}{Model parameters} & Background & \multicolumn{4}{c}{Phase factors} & \multicolumn{1}{c}{Electric moment} & \multicolumn{1}{c}{Sum rule} \\
		\cmidrule{1-1}\cmidrule{2-2}\cmidrule{3-6}\cmidrule{7-7}\cmidrule{8-8}
		$H^{(4)}_{1b}\oplus H^{(4)}_{2c}(t_x,t_y,t)$ & $\phi_0$ & $\phi_{pp}$ & $\phi_{po}$ & $\phi_{op}$ & $\phi_{oo}$ & $Q^{(2)}_c$ & $\sum (-1)^{ab} \phi_{ab}$\\
		\hline
		\multirow{4}{*}{$(0,0,0.1)$} 	
		& -0.25 & 0.33354(0) & -0.33994(1) & -0.07653(1) & 0.24992(1) & \multirow{4}{*}{0.25000(0)} & -0.00007(1) \\
		& 0.5 & 0.14434(1) & -0.49995(1) & -0.10573(0) & 0.24997(0) &  & -0.00002(0) \\
		& 0.25 & 0.43112(2) & -0.15944(1) & -0.15944(1) & 0.24999(0) &  & 0.00000(0) \\
		& 0 & -0.24983(2) & -0.49992(1) & -0.49992(1) & 0.24999(0) &  & 0.00000(0) \\
		\hline
		\multirow{4}{*}{$(0.1,0.1,0.1)$} 	
		& -0.25 & 0.374(2) & -0.307(1) & -0.0683(3) & 0.24992(1) & \multirow{4}{*}{0.25000(0)} & -0.00007(1) \\
		& 0.5 & 0.1592(5) & 0.49999(0) & -0.0908(5) & 0.24997(0) &  & -0.00002(0) \\
		& 0.25 & 0.489(2) & -0.131(1) & -0.131(1) & 0.24999(0) &  & 0.00000(0) \\
		& 0 & -0.24984(2) & -0.49992(1) & -0.49992(1) & 0.24999(0) &  & 0.00000(0) \\
		\hline
		\multirow{4}{*}{$(0.01,0.1,0.1)$} 	
		& -0.25 & 0.33455(4) & -0.33916(4) & -0.07630(0) & 0.24992(1) & \multirow{4}{*}{0.25000(0)} & -0.00007(1) \\
		& 0.5 & 0.14470(1) & 0.49999(0) & -0.10530(1) & 0.24997(0) &  & -0.00002(0) \\
		& 0.25 & 0.43274(8) & -0.15863(4) & -0.15863(4) & 0.24999(0) &  & 0.00000(0) \\
		& 0 & -0.24983(2) & -0.49992(1) & -0.49992(1) & 0.24999(0) &  & 0.00000(0) \\
		\end{tabular}
	\end{ruledtabular}
	\caption{
		The phase factors $\phi_{ab}$ and the corner charge $Q_c^{(2)}$	are computed for $H^{(4)}_{1b}\oplus H^{(4)}_{2c}$ in Eq.~\eqref{eq:ham-C4-1b-2c} with various parameters. On each $\phi_0$, all observables are obtained by extrapolating them as a function of $1/\sqrt{L_x L_y}$. The phase factors $\phi_{ab}$ converge to different values on each $\phi_0$, so we separate the convergence values of $\phi_{ab}$ by each value of $\phi_0$. Each extrapolation uses a sequence of four system sizes with the interval $(\Delta L_x, \Delta L_y)=(4,4)$, and each sequence starts from $(L_x,L_y)=(18,16),(18,17),(18,18),$ and $(19,19)$, which correspond to $\phi_0=-0.25, 0.5, 0.25$, and $0$, respectively. Here, we use the same extrapolation procedure as in Table.~\ref{table:quad-insulator}. The corner charge agrees with $\phi_{oo}$ and the sum rule Eq.~\eqref{eq:phi_quad_sum_rule} is satisfied up to small errors.
	}
	\label{table:C41b+2c-insulator}
\end{table*}

\section{Conclusion}\label{sec:conclusion}
In this paper, we have presented the bulk-boundary correspondence for the bulk quadrupole moment which is expressed in terms of the phase factors of the expectation values of the many-body operators with respect to the ground states under various boundary conditions. Our bulk-boundary correspondence is given by the cancellation between the four gauge-invariant phase factors, which can be computed in fully interacting systems. We also have proposed that one of them is expected to be identified with the corner charge. On these, we have numerically observed that when the band insulator without the hybrid Wannier value at $0.5$, each phase factor corresponds to a physical observable including the bulk multipole moment and edge-localized polarizations. Whether the same correspondence holds with the hybrid Wannier value at $0.5$ has not been fully understood yet. Through extensive numerical computations on band insulators, we have found that our bulk-boundary correspondence and the identification of one of the phase factors with the corner charge hold up to small errors of $O(10^{-4})$, which might be originated from the finite-size effect of the numerics. Furthermore, we have also found that another two of the phase factors have values similar to the edge-localized polarizations defined in Ref.~\onlinecite{Benalcazar2017} with differences at most $O(10^{-3})$, which supports our observations. 

Let us conclude by making remarks on possible future directions of our work. It would be interesting to prove the sum rules at least in the case of band insulators. This would also extend our knowledge of bulk multipole moments in solids. Since our formulation works for disordered and interacting systems, one could also test our sum rules for those cases. Finally, finding connections between previous works~\cite{PhysRevResearch.2.043012, PhysRevB.103.035147, ying2021change} and our sum rule would be interesting.

\acknowledgements
WL and GYC acknowledge the support of the National Research Foundation of Korea (NRF) funded by the Korean Government (Grant No. 2020R1C1C1006048 and 2020R1A4A3079707), as well as Grant No. IBS-R014-D1. This work is also supported by the Air Force Office of Scientific Research under Award No. FA2386-20-1-4029. BK is supported by KIAS individual Grant PG069402 at Korea Institute for Advanced Study and the National Research Foundation of Korea (NRF) grant funded by the Korea government (MSIT) (No. 2020R1F1A1075569). GYC acknowledges financial support from Samsung Science and Technology Foundation under Project Number SSTF-BA2002-05.

\appendix 

\section{Classical Multipolar Sum Rules}\label{appendix:classical_multipole_sum_rule}
In this appendix, we derive classical dipolar and quadrupolar sum rules based on classical electrostatics. The derivations closely follow the construction presented in Ref.~\onlinecite{Benalcazar2017}. Here, we consider continuum systems for classical systems.

\subsection{Classical Dipolar Sum Rule}\label{appendix:classical_dipole_sum_rule}
\begin{figure}[t]
	\centering\includegraphics[width=\columnwidth]{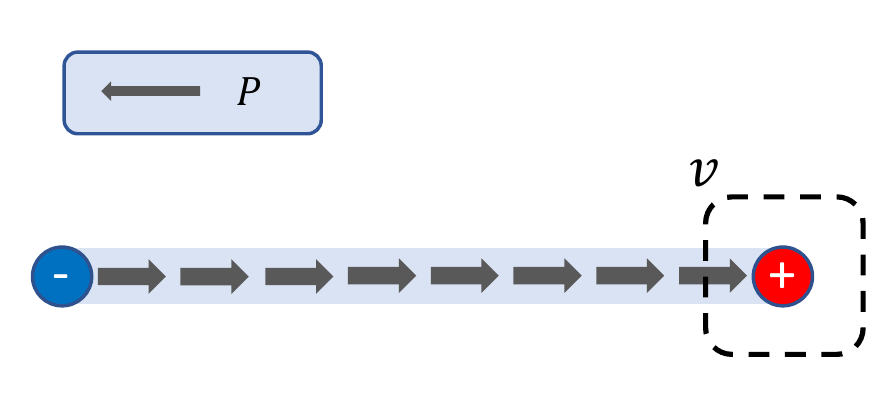}
	\caption{A schematics of the polarization in a one-dimensional system. Here, gray arrows represent the bulk polarization. The bulk polarization $P$ accumulates charges at each boundary with the same amount but different signs.}
	\label{Fig:pol}
\end{figure}

The macroscopic polarization $\vec{\mathcal{P}}(\vec{R})$ is defined in the classical electrostatics as the averaged first moment of the charge density over a region $v(\vec{R})$, which is small compared with the whole system,
\begin{equation}
	\vec{\mathcal{P}}(\vec{R}) = \frac{1}{|v(\vec{R})|}\int_{v(\vec{R})} d^3r \rho(\vec{r}+\vec{R})\vec{r},
\end{equation}
where $\rho(\vec{r})$ is the volume charge density and $|v(\vec{R})|$ is the volume of the region $v(\vec{R})$. It is well-known that a macroscopic system with the bulk polarization $\vec{\mathcal{P}}$ induces the charge density $\rho$ via
\begin{equation}\label{eq:rho_p}
	\rho = -\grad\cdot\vec{\mathcal{P}}.
\end{equation}
Suppose that the system has a boundary to the vacuum so that the polarization drops on the boundary, then the boundary charge $\mathcal{Q}_c$ is accumulated on the boundary:
\begin{equation}\label{eq:classical_p_Qc}
	\mathcal{Q}_c = -\int_v \grad\cdot\vec{\mathcal{P}} dv = -\oint_{\partial v} \vec{\mathcal{P}}\cdot d\vec{s},
\end{equation}
where $\partial v$ is the boundary of a volume $v$ which encircles the boundary of the system as shown in Fig.~\ref{Fig:pol}, and $\vec{s}$ is the surface vector normal to $\partial v$. If the system is one-dimensional and has a uniform bulk polarization density $\mathcal{P}$ along the $x$-direction, then Eq.~\eqref{eq:classical_p_Qc} becomes
\begin{equation}
	\mathcal{Q}_c = \mathcal{P},
\end{equation}
which is depicted in Fig. \ref{Fig:pol}, where the circle containing the plus symbol being the boundary charge.

\subsection{Classical Quadrupolar Sum Rule}\label{appendix:classical_quadrupole_sum_rule}
\begin{figure}[t]
	\centering\includegraphics[width=\columnwidth]{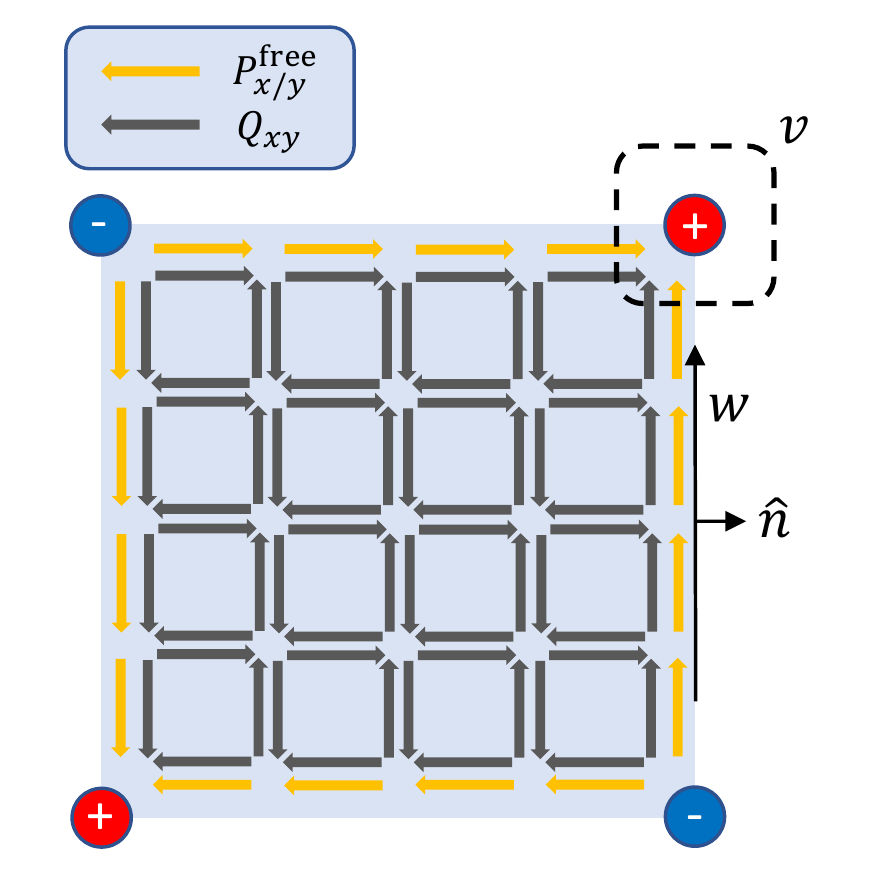}
	\caption{Schematic understanding of the bulk quadrupole moment and the free edge polarization.
			 Here, gray arrows and orange arrows represent the quadrupole moment and the free edge polarization, respectively.
			 Both of them accumulate charges at each corner, and the total amount of charge at each corner is fixed by
			 the charge conservation law Eq.~\eqref{eq:classical_qxy_Qc}.}
	\label{Fig:quadpol}
\end{figure}

The macroscopic quadrupole moment $\mathcal{Q}_{ij}(\vec{R})$ is defined in the classical electrostatics as the averaged second moment of the charge density over $v(\vec{R})$,
\begin{equation}
	\mathcal{Q}_{ij}(\vec{R}) = \frac{1}{|v(\vec{R})|}\int_{v(\vec{R})} d^3r \rho(\vec{r}+\vec{R})r_i r_j,
\end{equation}
where $\rho(\vec{r})$ is the volume charge density.
We first consider a classical system with a uniform bulk quadrupole moment $\mathcal{Q}_{ij}$. We assume that the system has a vanishing bulk polarization, which follows when the system has the inversion symmetry. Then the quadrupole moment $\mathcal{Q}_{ij}$ induces the charge density $\rho$ via
\begin{equation}\label{eq:rho_qxy}
	\rho = \frac{1}{2}\sum_{i,j}\partial_i\partial_j \mathcal{Q}_{ij}.
\end{equation}
If we consider a rectangular geometry as in Fig.~\ref{Fig:quadpol}, then the corner charge $\mathcal{Q}_c$ accumulated on a corner of a macroscopic two-dimensional system due to quadrupole moment $\mathcal{Q}_{ij}$ drop is given by
\begin{equation}\label{eq:classical_qxy_Qc}
	\mathcal{Q}_c = \frac{1}{2}\sum_{i,j}\int_v \partial_i\partial_j \mathcal{Q}_{ij} dv 
	= \frac{1}{2}\sum_{i,j}\oint_{\partial v} \partial_i \mathcal{Q}_{ij}n_j ds,
\end{equation}
where $n_i$ is the $i$-th component of the unit vector normal to $\partial v$,
$v$ is a volume envelops the corner and $\partial v$ is the boundary of $v$, which are illustrated in Fig.~\ref{Fig:quadpol}. For a rectangular system, Eq.~\eqref{eq:classical_qxy_Qc} becomes
\begin{equation}
	\mathcal{Q}_c = \mathcal{Q}_{xy},
\end{equation}
where we used the symmetry of the quadrupole moment, $\mathcal{Q}_{xy}=\mathcal{Q}_{yx}$. 
In addition to inducing corner charge, the quadrupole moment also induces the polarizations along the edges, which we call the edge-localized polarizations. To this end, we first consider the geometry given in Fig.~\ref{Fig:quadpol}. The line charge density $\sigma_w$ at the edge $w$ in Fig.~\ref{Fig:quadpol} is given as
\begin{equation}
	\sigma_w = -\sum_{i,j} \partial_j (n_i \mathcal{Q}_{ij}),
\end{equation}
where $n_i$ is the normal vector of the edge in Fig.~\ref{Fig:quadpol}.
Since the charge density is given as the divergence of $\sum_i n_i \mathcal{Q}_{ij}$, 
this can be considered as the polarization induced by the quadrupole moment localized on the edge $w$.
We will call this polarization as $\mathcal{P}^{\text{quad}}_n$.
For a rectangular system, the quadrupole induced polarizations at 
the $y$-boundaries $\mathcal{P}^{\text{quad}}_x$ and the $x$-boundaries $\mathcal{P}^{\text{quad}}_y$ are given by
\begin{equation}
	\mathcal{P}^{\text{quad}}_x = \mathcal{P}^{\text{quad}}_y = \mathcal{Q}_{xy}.
\end{equation}
On the other hand, one can dress one dimensional systems having edge polarizations $\mathcal{P}^{\text{free}}_{x}$ and $\mathcal{P}^{\text{free}}_{y}$ along the boundaries while respecting the inversion symmetry, where $\mathcal{P}^{\text{free}}_{x}$ and $\mathcal{P}^{\text{free}}_{y}$ do not come from the quadrupole moment. Note that $\mathcal{P}^{\text{free}}_{x/y}$ also accumulates the charge $\mathcal{Q}^\textrm{free}_c$ at the corner due to the dipolar sum rule 
\begin{equation}
	\mathcal{Q}^\textrm{free}_c = \mathcal{P}^{\text{free}}_x + \mathcal{P}^{\text{free}}_y.
\end{equation}
Thus, if we define the edge-localized polarization $\mathcal{P}^{\text{edge}}_{x/y}$ in classical systems as
\begin{equation}
	\mathcal{P}^{\text{edge}}_{x/y} \equiv \mathcal{P}^{\text{quad}}_{x/y} + \mathcal{P}^{\text{free}}_{x/y},
\end{equation}
then the corner charge is given by
\begin{equation}\label{eq:classical_quad_sum_rule}
	\mathcal{Q}_c = -\mathcal{Q}_{xy} + \mathcal{P}^{\text{edge}}_{x} + \mathcal{P}^{\text{edge}}_{y}.
\end{equation}
We call Eq.~\eqref{eq:classical_quad_sum_rule} the quadrupolar sum rule in classical systems, which is summarized in Fig.~\ref{Fig:quadpol}. 

\section{$\hat{U}_1$ operator and the corner charge}\label{appendix:U1}
In this appendix, we show that for band insulators in one dimension, the phase factor of the $\hat{U}_1$ expectation value with respect to the ground state in the open boundary condition converges to the boundary charge in the thermodynamic limit. 

We prove the claim by constructing bulk and boundary Wannier functions. The bulk Wannier functions are identical to the Wannier functions in the periodic boundary condition and the boundary Wannier functions are localized near the left and right boundaries. The phase factor of $\hat{U}_1$ expectation value is determined by the bulk Wannier functions while boundary Wannier functions do not contribute, as in the original bulk-boundary correspondence for the polarization~\cite{PhysRevB.48.4442, Rhim2017}. To avoid any ambiguity, we assume that no mode exists at the chemical potential, which can always be done by either introducing a (small) symmetry breaking term splitting the zero modes or tuning the chemical potential slightly.

In the open boundary condition, the position operator $\hat{x}$ is well-defined, which we choose to take values in $\{1, 2, \cdots, L\}$, and thus we can diagonalize the position operator in the subspace of occupied single-particle orbitals $P_\textrm{occ}$. We would like to show that the eigenvectors of $P_\textrm{occ} ( \hat{x} ) P_\textrm{occ}$ are exactly the bulk and boundary Wannier functions with desirable properties.

We first recall the results in Refs.~\onlinecite{PhysRevB.26.4269, Alexandradinata2014}, which state that the Wannier functions $\{ \vert w_{\alpha, R} \rangle \}$, where $\alpha = 1, \cdots, n_e$ with $n_e$ being the electron filling and $R \in \mathbb{Z}$ labels the position at which the Wannier function is localized, are the eigenvectors of $\hat{x}$ operator in the subspace of the occupied single-particle orbitals of the {\it infinite} open system with the corresponding eigenvalues $R + \nu_\alpha$. At the same time, $\{ \vert w_{\alpha, R} \rangle \}$ are the eigenvectors of $e^{\frac{2\pi i \hat{x}}{L}}$ in the subspace of occupied single-particle orbitals of a periodic system with the corresponding eigenvalues $e^{i2 \pi (R+\nu_\alpha)/L}$, when the system size $L$ is sufficiently large. Moreover, the Wannier functions $\{ \vert w_{\alpha, R} \rangle \}$ are exponentially localized and satisfy the usual property $\langle x \vert w_{\alpha, R+1} \rangle = \langle x-1 \vert w_{\alpha, R} \rangle$. 

Let us take a positive integer $\delta$ which is much larger than the localization lengths of $\{ \vert w_{\alpha, R} \rangle \}$. Using the exponentially localized nature, $\{ \vert w_{\alpha, R} \rangle \}_{R =\delta+1, \delta_2, \cdots, L-\delta}$ are the claimed bulk Wannier functions, i.e., eigenvectors of $P_\textrm{occ} ( \hat{x} ) P_\textrm{occ}$ localized on the bulk, where $P_\textrm{occ}$ projects to the subspace of occupied single-particle orbitals of a finite open system. Also, for a sufficiently large $L$, the bulk Wannier functions $\{ \vert w_{\alpha, R} \rangle \}_{R =\delta+1, \delta_2, \cdots, L-\delta}$ are the eigenvectors of $P_\textrm{occ} ( e^{\frac{2\pi i \hat{x}}{L}} ) P_\textrm{occ}$ with the corresponding eigenvalues $e^{i2 \pi (R+\nu_\alpha)/L}$. The remaining eigenvectors of $P_\textrm{occ} ( \hat{x} ) P_\textrm{occ}$ are localized near the left and right boundaries, hence we call them the boundary Wannier functions.

We now evaluate the expectation value of $\hat{U}_1$ using the bulk and boundary Wannier functions. Since the ground state of a band insulator is given by the slater determinant of occupied single-particle orbitals, which we choose to be the bulk and boundary Wannier functions, $\langle \hat{U}_1 \rangle$ can be expressed as
\begin{equation}
	\langle \hat{U}_1 \rangle = e^{-(L+1) n_e \pi i} \det \left[ \left(
	\begin{array}{ccc}
	\mathcal{L} & 0 & 0 \\
	0 & \mathcal{D} & 0 \\
	0 & 0 & \mathcal{R}
	\end{array}
	\right) \right],
\end{equation}
where $e^{-(L+1) n_e \pi i}$ is the phase factor from background ions, $\mathcal{L}$ and $\mathcal{R}$ are $n_e \delta$-by-$n_e \delta$ matrices, and 
\begin{equation}
	\mathcal{D} = \textrm{diag}( e^{2\pi i \frac{\delta+1+\nu_1}{L}} , \cdots, e^{2\pi i \frac{L-\delta+\nu_{N_\textrm{oc}}}{L}} )
\end{equation}
is a diagnal matrix. By expanding the diagonal part in the determinant, we get
\begin{align}
	\langle U_1 \rangle &= e^{-(L+1) n_e \pi i} e^{\frac{2\pi i}{L} \big( \frac{(L+1)(L-2\delta)}{2} n_e + L \sum_{\alpha=1}^{n_e} \nu_\alpha \big)} \nonumber \\
	& \quad \times \det (\mathcal{L}) \det (\mathcal{R}) \nonumber \\
	&= e^{2\pi i \frac{L+1}{L} (-\delta)} e^{2\pi i\sum_{\alpha=1}^{n_e} \nu_\alpha} \det (\mathcal{L}) \det (\mathcal{R}) \nonumber \\
	&= e^{-2\pi i \frac{\delta}{L}} e^{2\pi i\sum_{\alpha=1}^{n_e} \nu_\alpha} \det (\mathcal{L}) \det (\mathcal{R}) ,
\end{align}
where we used the fact that $\delta$ is an integer in the last equality. Note that $\mathcal{L}$ ($\mathcal{R}$) is the matrix representation of $e^{\frac{2\pi i \hat{x}}{L}}$ with respect to the left (right) Wannier functions which are localized within the range $\delta$.  Thus, in the thermodynamic limit $\hat{U}_1 = I + O(\frac{1}{L})$ on the left (right) Wannier functions and hence $\det(\mathcal{L}) = 1 + O(\frac{1}{L})$ ($\det(\mathcal{R}) = 1 + O(\frac{1}{L})$). Thefore,
\begin{equation}
	\lim_{L \to \infty} \frac{1}{2\pi} \textrm{Im} \Big[ \log \big( \langle U_1 \rangle \big) \Big] = \sum_{\alpha=1}^{N_\textrm{occ}} \nu_\alpha = Q_c^{(1)} ,
\end{equation}
where the last equality follows from the original bulk-boundary correspondence~\cite{PhysRevB.48.4442, Rhim2017}. This completes the proof that $\hat{U}_1$ expectation value reproduces the boundary charge for band insulators in the open boundary condition.

\section{Sum rule in $C_3$-symmetric insulator} \label{appendix:C3-model}
As a demonstration that the sum rule Eq.~\eqref{eq:phi_quad_sum_rule} works for insulators having other types of symmetry besides $C_4$ symmetry, we consider a $C_3$-symmetric insulator~\cite{Benalcazar2019} and numerically confirm the sum rule on this model.

\subsection{$C_3$-symmetric insulator: $H^{(3)}_{2b}\oplus H^{(3)}_{2c}$}
The tight-binding hamiltonian in the momentum space of the $C_3$-symmetric insulator $H^{(3)}_{2b}\oplus H^{(3)}_{2c}$ is given by~\cite{Benalcazar2019}
\begin{equation}\label{eq:ham-C3-2b-2c}
	H^{(3)}_{2b}\oplus H^{(3)}_{2c}(\boldsymbol{k}) = 
	\begin{bmatrix}
		H^{(3)}_{2b} & \gamma^{(3)}(t) \\
		\gamma^{(3)}(t)^\dagger & H^{(3)}_{2c}
	\end{bmatrix} ,
\end{equation}
where
\begin{equation}
	H^{(3)}_{2b}(\boldsymbol{k}) = 
	\begin{bmatrix}
		0 & t_0+e^{i\boldsymbol{k}\cdot\boldsymbol{a_2}} & t_0+e^{-i\boldsymbol{k}\cdot\boldsymbol{a_3}}\\
		t_0+e^{-i\boldsymbol{k}\cdot\boldsymbol{a_2}} & 0 & t_0+e^{-i\boldsymbol{k}\cdot\boldsymbol{a_1}}\\
		t_0+e^{i\boldsymbol{k}\cdot\boldsymbol{a_3}} & t_0+e^{i\boldsymbol{k}\cdot\boldsymbol{a_1}} & 0
	\end{bmatrix} ,
\end{equation}
\begin{equation}
	H^{(3)}_{2c}(\boldsymbol{k}) = 
	\begin{bmatrix}
		0 & t_0+e^{i\boldsymbol{k}\cdot\boldsymbol{a_1}} & t_0+e^{i\boldsymbol{k}\cdot\boldsymbol{a_2}}\\
		t_0+e^{-i\boldsymbol{k}\cdot\boldsymbol{a_1}} & 0 & t_0+e^{-i\boldsymbol{k}\cdot\boldsymbol{a_3}}\\
		t_0+e^{-i\boldsymbol{k}\cdot\boldsymbol{a_2}} & t_0+e^{i\boldsymbol{k}\cdot\boldsymbol{a_3}} & 0
	\end{bmatrix} ,
\end{equation}
and
\begin{equation}\label{eq:C3-interlayer}
	\gamma^{(3)}(t) = 
	\begin{bmatrix}
		t & t & 0\\
		0 & t & t\\
		t & t & 0
	\end{bmatrix}.
\end{equation}
The $2/3$-filled ground states of $H^{(3)}_{2b}$ and $H^{(3)}_{2c}$ have non-vanishing polarizations $1/3$ and $2/3$, respectively, so the stacked model $H^{(3)}_{2b}\oplus H^{(3)}_{2c}$ at $2/3$-filling with the inter-layer hopping term $\gamma^{(3)}(t)$ has vanishing bulk polarization and respects $C_3$-symmetry. 

We will consider the stacked model Eq.~\eqref{eq:ham-C3-2b-2c} on the square lattice. To do so, we use the coordinates system $(x,y)$ with two basis vectors $\vec{a}_1=(1,0)$ and $\vec{a}_2=(1/2,\sqrt{3}/2)$, i.e., the coordinates $(x,y)$ refer to as the vector of $x\vec{a}_1+y\vec{a}_2$. All the phase factors $\phi_{ab}$ defined in Eq.~\eqref{eq:U2_phase} and appears in Table~\ref{table:C3-2b-2c} are computed using these coordinates. We find that the stacked model on the square lattice at $2/3$-filling has corner charges $1/3$ and $-1/3$ well localized near the corners.

The numerical results are summarized in Table~\ref{table:C3-2b-2c}. We see that the corner charge $Q^{(2)}_c$ agrees with $\phi_{oo}$ and the sum rule Eq.~\eqref{eq:phi_quad_sum_rule} is satisfied up to small errors of $O(10^{-5})$.

\begin{table*}
	\footnotesize
	\centering
	\begin{ruledtabular}
	\renewcommand{\arraystretch}{1.4}
	\begin{tabular}{rrrrrrr}
		\multicolumn{1}{c}{Model parameters} & \multicolumn{4}{c}{Phase factors} & \multicolumn{1}{c}{Electric moment} & \multicolumn{1}{c}{Sum rule}\\
		\cmidrule{1-1}\cmidrule{2-5}\cmidrule{6-6}\cmidrule{7-7}
		$H^{(3)}_{2b}\oplus H^{(3)}_{2c}(t_0,t)$ & $\phi_{pp}$ & $\phi_{po}$ & $\phi_{op}$ & $\phi_{oo}$ & $Q^{(2)}_c$ & $\sum (-1)^{ab} \phi_{ab}$\\
		\hline
        $(0.001,0.001)$ & -0.33340(0) & -0.33333(0) & -0.33333(0) & -0.33331(0) & -0.33333(0) & -0.00005(0) \\
        $(0.1,0.001)$ & -0.33340(0) & -0.33335(0) & -0.33335(0) & -0.33334(0) & -0.33337(0) & -0.00005(0) \\
        $(0.1,0.1)$ & -0.33377(0) & -0.33441(0) & -0.33441(0) & -0.33510(0) & -0.33513(0) & -0.00004(0) \\
        $(0.2,0.1)$ & -0.33380(0) & -0.33504(0) & -0.33504(0) & -0.33631(0) & -0.33634(0) & -0.00004(0) \\
		\end{tabular}
	\end{ruledtabular}
	\caption{
		The phase factors $\phi_{ab}$ and the corner charge $Q_c^{(2)}$	are computed for $H^{(3)}_{2b}\oplus H^{(3)}_{2c}(t_0,t)$ in Eq.~\eqref{eq:ham-C3-2b-2c} on a square lattice with various parameters. The details about the lattice and filling are discussed in the paragraph below Eq.~\eqref{eq:C3-interlayer}. Here we use the same extrapolation procedure as in Table.~\ref{table:quad-insulator}. The corner charge agrees with $\phi_{oo}$ and the sum rule Eq.~\eqref{eq:phi_quad_sum_rule} is satisfied up to small errors.
	}
	\label{table:C3-2b-2c}
\end{table*}

\section{Branch cut dependence of the HWF-based edge-localized polarization}\label{appendix:edge-pol}
In this section, we discuss the branch cut dependence in the hybrid Wannier values of the HWF-based edge-localized polarization $\tilde{P}^{\textrm{edge}}_{x/y}$ defined in Ref.~\onlinecite{Benalcazar2017} and Eq.~\eqref{eq:edgepol}. To be self-contained, we first rewrite the definition of the edge-localized polarization,
\begin{equation}\label{eq:edgepol_appendix}
	\tilde{P}^{\textrm{edge}}_x = \sum_j \sum_{y=1}^{L_y/2}\nu^j \rho^j(y)\quad\text{(mod 1)},
\end{equation}
where $\rho^j(y)$ and $e^{2\pi i\nu_j}$ are the density and the $j$-th eigenvalue of the hybrid Wilson loop $\mathcal{W}_{k_x}$ along the $x$-direction, respectively. We will call $\nu^j$ the hybrid Wannier value.

The edge-localized polarization $\tilde{P}^{\textrm{edge}}_{x}$ can be well defined after one fixes the branch cut of the hybrid Wannier value since the hybrid Wannier value is defined by a phase angle, which has the modulo $2\pi$ ambiguity. Unless fixing the branch cut, each $\nu^j$ can be shifted by an integer $n^j=\pm 1$, which results in shifting $\tilde{P}^{\textrm{edge}}_{x}$ as
\begin{equation}
	\Delta \tilde{P}^{\textrm{edge}}_x=\sum_j \sum_{y=1}^{L_y/2} n^j \rho^j(y).
\end{equation}
Since $\sum_{y=1}^{L_y/2} \rho^j(y)$ is not quantized in general, $\Delta \tilde{P}^{\textrm{edge}}_x$ may also not be quantized if $n^j$ is not the same for all $j$. Consequently, the edge-localized polarization is ambiguous before fixing the branch cut even if we take the modulo one equivalence on both sides of Eq.~\eqref{eq:edgepol_appendix}. Here we fix the branch cut by fixing the range of the hybrid-Wannier value as $\nu\in(-0.5,0.5]$. This choice is made for our numerical results to be consistent with that of Ref.~\onlinecite{Benalcazar2017}. Note that another equally valid choice of the range is $\nu\in[-0.5,0.5)$, which includes $-0.5$ instead of $+0.5$.

However, these possibilities in ranges again cause a problem when a hybrid Wannier value $\nu^j$ is at the branch cut value since $\nu^j$ can be either $+0.5$ or $-0.5$ depending on the choice of the range. In this case, the edge-localized polarization is invariant under the choice of the range only when $\sum_{y=1}^{L_y/2} \rho^j(y)$ is quantized, i.e., the HWF is localized one of the two half systems. The situation gets worse when more than two hybrid Wannier values $\nu^1, \nu^2, ..., \nu^n$ are degenerated at the brach cut since for all possible unitary transformations $\tilde{\psi}^i(y)=\sum_{j=1}^nu_{ij}\psi^j(y)$ of the Wannier functions $\{\psi^1, \psi^2, ..., \psi^n\}$, $\sum_{y=1}^{L_y/2} \abs{\tilde{\psi}^i(y)}^2$ should be quantized for the edge-localized polarization not to depend sensitively on the choice of the branch cut. To avoid this difficulty, we computed $\tilde{P}^\textrm{edge}_{x/y}$ only for the models without the hybrid Wannier value at $0.5$ in the main text.

\bibliography{ref}

\end{document}